\documentclass[preprint2]{aastex}
\usepackage{natbib}
\slugcomment{To appear in The Astronomical Journal}
\received{15.11.2001}
\revised{11.1.2002}
\accepted{17.1.2002}
\shorttitle{Dust Beyond Jupiter}
\shortauthors{Landgraf et al.}
\begin{document}
\title{Origins of Solar System Dust Beyond Jupiter}
\author{M. Landgraf\altaffilmark{1,3}, J.-C. Liou\altaffilmark{2},
H. A. Zook\altaffilmark{3}, and E. Gr\"un\altaffilmark{4}}
\altaffiltext{1}{ESA/ESOC, 64293 Darmstadt}
\email{Markus.Landgraf@esa.int}
\altaffiltext{2}{Lockheed Martin Space Operations, Houston, TX}
\altaffiltext{3}{NASA/JSC, Houston, TX}
\altaffiltext{4}{MPI-K, 69029 Heidelberg}

\begin{abstract}
The measurements of cosmic interplanetary dust by the instruments on
board the Pioneer 10 and 11 spacecraft contain the dynamical signature
of dust generated by Edgeworth-Kuiper Belt objects, as well as short
period Oort Cloud comets and short period Jupiter family comets. While
the dust concentration detected between Jupiter and Saturn is mainly
due to the cometary components, the dust outside Saturn's orbit is
dominated by grains originating from the Edgeworth-Kuiper Belt. In
order to sustain a dust concentration that accounts for the Pioneer
measurements, short period external Jupiter family comets, on orbits
similar to comet 29P/Schwassmann-Wachmann-1, have to produce $8\times
10^4\:{\rm g}\:{\rm s}^{-1}$ of dust grains with sizes between $0.01$
and $6\:{\rm mm}$. A sustained production rate of $3\times 10^5\:{\rm
g}\:{\rm s}^{-1}$ has to be provided by short period Oort cloud comets
on 1P/Halley-like orbits. The comets can not, however, account for the
dust flux measured outside Saturn's orbit. The measurements there can
only be explained by a generation of dust grains in the
Edgeworth-Kuiper belt by mutual collisions of the source objects and
by impacts of interstellar dust grains onto the objects'
surfaces. These processes have to release in total $5\times 10^7\:{\rm
g}\:{\rm s}^{-1}$ of dust from the Edgeworth Kuiper belt objects in
order to account for the amount of dust found by Pioneer beyond
Saturn, making the Edgeworth-Kuiper disk the brightest extended
feature of the Solar System when observed from afar.
\end{abstract}


\section{Introduction}
Our Solar System as well as other planetary systems is filled with
small solid particles, either interstellar survivors of the formation
process, or fragments of larger bodies like asteroids, comets, moons,
or planets. Commonly referred to as interplanetary dust, these
particles carry information about their sources, not only by their
chemical signature \citep{brownlee85,kissel86}, but also by the size
and shape of their orbits around the Sun. The particles' chemistry as
well as their orbit can best be gauged in situ, that is by dust
detectors on board interplanetary spacecraft. While the accretion of
interplanetary dust particles by the Earth's atmosphere allows their
mineralogical, chemical, and isotopic analysis in ground-based
laboratories after their collection by high-flying aircraft,
information on their orbit around the Sun is lost after the
atmospheric entry. The orbital properties of Solar System dust inside
Jupiter's orbit has been extensively studied by in situ measurements
\citep{mcdonnell75,gruen77,gruen95b,gruen95c,brownlee97}. From these
measurements Jupiter family short period comets and asteroids have
been identified as the dominant dust sources
\citep{liou95,dermott92}. In the grain size regime below $1\:{\rm \mu
m}$ a high abundance of interstellar grains was found
\citep{gruen93}. While interstellar impactors can easily be
distinguished from detections caused by solar system dust, it is still
unclear what the relative contribution of the various interplanetary
sources is. Besides this uncertainty, the large number of in situ
measurements taken inside Jupiter's orbit led to a consistent picture
of the extend and distribution solar system dust cloud there. The
situation beyond Jupiter's orbit is however vastly different. So far
the only in situ dust detectors ever to fly beyond Jupiter are the
dust experiments on board the Pioneer 10 and 11 spacecraft
\citep{humes80}\footnote{Since 31 December 2000, the Cassini spacecraft
is outside Jupiter's orbit on its way to its final destination
Saturn.}. Measurements of the plasma instruments on board Voyager 1
and 2 seem to indicate a high concentration of micron-sized particles
out to $50\:{\rm AU}$ \citep{gurnett97}. The Voyager results are
however not conclusive because the plasma instruments have never been
calibrated to measure dust impacts. From the Pioneer 10 and 11
measurements \cite{humes80} found that, taken as an ensemble,
the particles have to have a constant spatial concentration as
function of the distance from the Sun and move on highly eccentric,
randomly oriented orbits. In this report we use the Pioneer 10 and 11
data to identify the source objects of the particles by modelling the
sources' signature in the Pioneer data, and comparing the measurements
with the result of the modelling.

\section{In situ Measurements beyond Jupiter by the Pioneer Missions}
\begin{figure*}
\centering
\epsscale{2}
\plotone{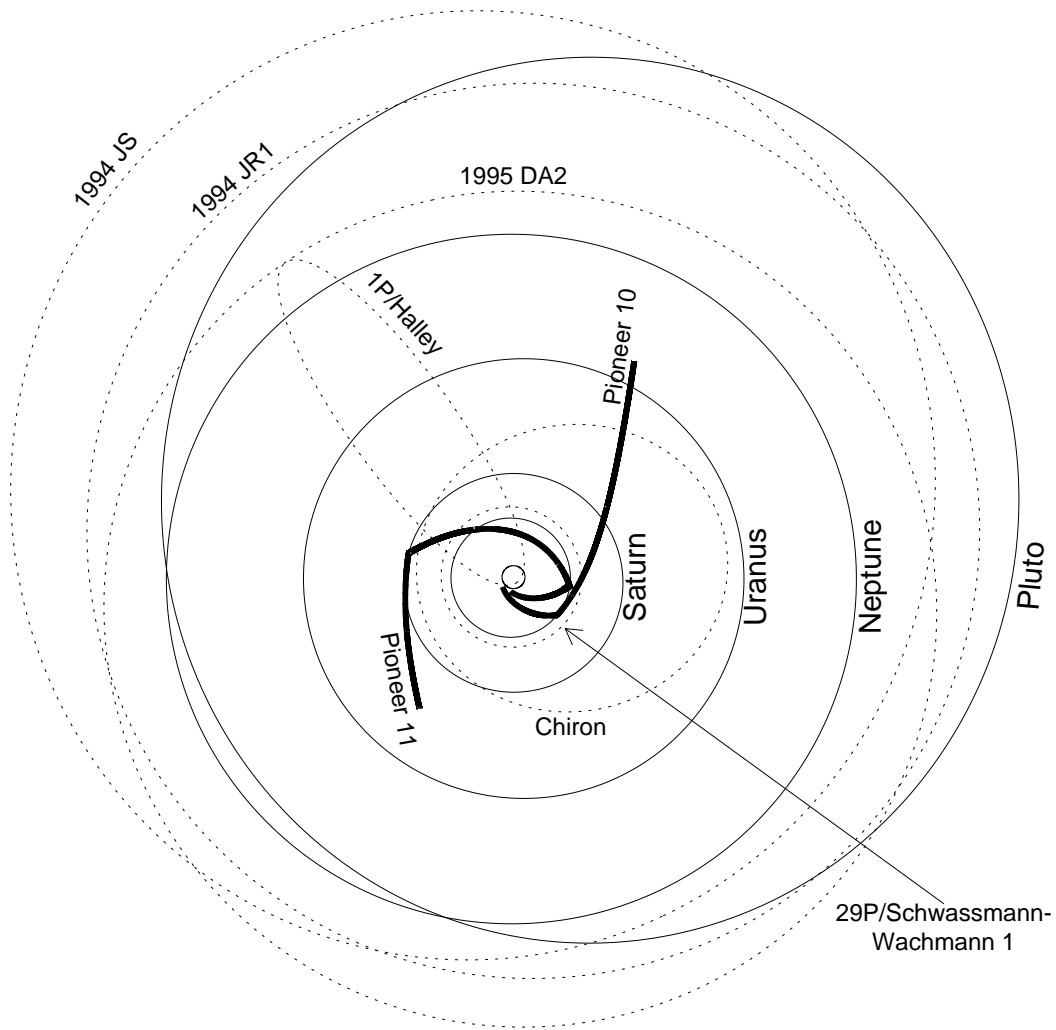}
\caption{\label{fig_overview}\it\small Overview of the orbits of the Pioneer
spacecraft (solid, thick) and potential dust source objects
(dotted). The orbits of the planets Earth, Jupiter, Saturn, Uranus,
Neptune, and Pluto are shown as the thin solid lines. As
representatives of the dust sources comets 1P/Halley and
29P/Schwassmann-Wachmann 1, the Centaur object 2060 Chiron, and the
transneptunian objects 1994 JS, 1994 JR1, and 1995 DA2 are shown.}
\end{figure*}

The Pioneer instruments consist of panels of 234 pressurised cells,
mounted on the back of the spacecraft's high gain antenna. The cells
are divided in two separate electronic channels for redundancy, 108
cells are connected to channel 0 and 126 cells are connected to
channel 1. Each cell has a cross section area of $2.45\times
10^{-3}\:{\rm m}^2$. The instruments register the time when a particle
penetrates the thin wall of the cell that encloses the pressurised
gas. Before the penetration the gas acts as an insulator between two
electrodes, and as it escapes into the vacuum of space, the electrodes
discharge and the resulting electrical signal is registered as a
penetration event. The sensitivity of the instrument, that is the
minimum impact mass and velocity that causes a penetration, is
determined by the thickness of the cell walls. On the Pioneer 10
experiment walls of $25\:{\rm \mu m}$ were used, and on Pioneer 11 the
cell walls were $50\:{\rm \mu m}$ thick. At a typical impact velocity
of $20\:{\rm km}\:{\rm s}^{-1}$, the Pioneer 10 cells are penetrated
by particles with an equivalent diameter larger than $10\:{\rm \mu
m}$, and the Pioneer 11 cells are penetrated by $21\:{\rm \mu m}$
particles \citep{humes74}\footnote{Assuming a grain mass density of
$1\:{\rm g}\:{\rm cm}^{-3}$.}. The surfaces of the Pioneer instruments
always point nearly opposite to the high gain antenna, away from the
Earth. Beyond Jupiter this means the instruments are oriented mainly
away from the Sun with an effective field of view of $1.6\pi\:{\rm
sr}$ ($240^\circ$ opening angle). The Pioneer 10 instrument took
measurements from the launch on 2 March 1972 until it failed on 10 May
1980 due to the low temperatures, $18\:{\rm AU}$ from the Sun (for the
geometry of the spacecraft trajectories see figure
\ref{fig_overview}). 
Pioneer 11 performed dust measurements from
launch on 5 April 1973 until it was switched off 25 September
1983. The Pioneer dust instruments successfully detected 225
penetrations altogether, however, they did not work flawlessly. On
Pioneer 10 one channel failed completely, and on Pioneer 11 an
unexplained discrepancy between the rate of penetrations measured by
both channels was observed. The flux measured by one channel of the
Pioneer 11 instrument is consistently higher than the flux measured by
the other. Because the angular sensitivity of both channels is
identical, this discrepancy can only be due to a malfunction of one of
the channels. Despite these inconsistencies we consider the Pioneer
dust data to be reliable for the following reasons: (a) the rate of
detected events increased sharply during the fly-bys of Jupiter and
Saturn which is not expected for random noise, and (b) the flux
densities measured by Pioneer 10 and 11 at $1\:{\rm AU}$ are in accord
with measurements by Explorer 23, an Earth orbiting spacecraft that
was equipped with similar instruments \citep{humes76}. The discrepancy
between the Pioneer 11 channels can be explained by either the loss of
cells on one of the channels during the launch of the spacecraft, or
by electronic noise in one of the channels. Figures \ref{fig_profile}a
and b show the interplanetary penetration flux\footnote{Penetrations
per unit area and time, sliding mean over $4$ penetration events,
penetrations during the fly-bys of the planets have been removed.} on
the Pioneer dust instruments as a function of time and distance from
the Sun. After the launch the dust flux measured by Pioneer 10 is
$2\times 10^{-5}\:{\rm m}^{-2}\:{\rm s}^{-2}$, continuously decreasing
with heliocentric distance to $3\times 10^{-6}\:{\rm m}^{-2}\:{\rm
s}^{-1}$ at Jupiter distance. After passing Jupiter's orbit, the flux
measured by Pioneer 10 stays almost constant. Due to the lower
abundance of large grains the fluxes measured by the less sensitive
Pioneer 11 instrument are smaller but they draw a similar picture:
decreasing flux from Earth to Jupiter, and an almost constant flux
outside Jupiter's orbit.

\section{Sources of Dust Beyond Jupiter}
What are the sources of the particles that penetrated the cells of the
Pioneer dust instruments? The interstellar dust stream that was
discovered by the dust instrument on board Ulysses causes an
approximately constant dust concentration around the Sun, which
potentially explains the constant penetration rate of the Pioneer
instrument. However, from the extrapolation of the flux-mass
distribution of interstellar dust measured by Ulysses to the Pioneer
10 threshold mass, it follows that less than $10^{-8}\:{\rm
m}^{-2}\:{\rm s}^{-1}$ interstellar penetrations of Pioneer 10 cells
can be expected \citep{landgraf00a}, less than one percent of the
measured flux. We are thus left with interplanetary particles as the
cause for the penetrations detected by the Pioneer dust
experiments. Since the abundance of interplanetary particles decreases
steeply with their size \citep{gruen85}, we can assume that the
penetrations were caused mainly by particles with sizes just above the
detection threshold of the instruments, i.e., with diameters in the
order of $10\:{\rm \mu m}$. Particles in this size regime move
approximately on Keplerian orbits, because their dynamics are
dominated by solar gravity. Over long time scales the orbits evolve
under Poynting-Robertson (PR) and solar wind drag. This drag force is
caused by the relativistic aberration of the sunlight and solar wind
particles \citep{burns79}. The effect of PR and solar wind drag is to
remove energy from the particle's orbit causing a slow inward directed
spiral motion of the particles. The aphelion of a source object of a
particle must therefore be equal or larger than the particle's
distance from the Sun. Consequently, the sources of the constant flux
of particles measured by Pioneer outside Jupiter must lie beyond
Jupiter's orbit. We distinguish 3 dynamic families that we consider as
potential dust sources: 1P/Halley-type comets (HTC, short period Oort
cloud comets), 29P/Schwassmann-Wachmann-1-type comets (SW1TC, short
period Jupiter family comets with perihelion close to Jupiter's
orbit), and Edgeworth-Kuiper belt objects (EKBOs). Both, 1P/Halley as
well as 29P/Schwassmann Wachmann 1 have been reported to be prolific
sources of dust \citep{kissel86,fulle92} as they disintegrate due to
solar heating. For EKBOs it is proposed that they release dust due to
mutual collisions \citep{backman95,stern96} and due to impacts by
interstellar particles \citep{yamamoto98}.  Another potential source
of dust outside Jupiter are Centaur objects that orbit the Sun between
Saturn and Uranus. They are however not considered strong sources,
because their number is too small to cause frequent collisions, and
dust particles released by them are likely to be ejected from the
solar system due to their highly eccentric orbits that cross the
orbits of several giant planets. They are also too far from the Sun to
exhibit a strong cometary activity \citep{brown98}. The dynamic
families of source objects described above are defined by their
interaction with the major planets. Comets are considered a HTC if
their perihelion is inside Jupiter's, their aphelion outside Neptune's
orbit, and their inclination between $160^\circ$ and
$180^\circ$. SW1TCs have their perihelion close to Jupiter's
orbit, an eccentricity below $0.1$, and an inclination below
$10^\circ$. Finally members of the EKBO family have perihelia beyond
Neptune, eccentricities below $0.1$, and inclinations below
$20^\circ$, which includes classical as well as scattered members of
the Edgeworth-Kuiper belt \citep{brown01}.

\section{Dust Distribution by Orbital Evolution}
\begin{figure*}
\centering
\epsscale{1.6}
\plotone{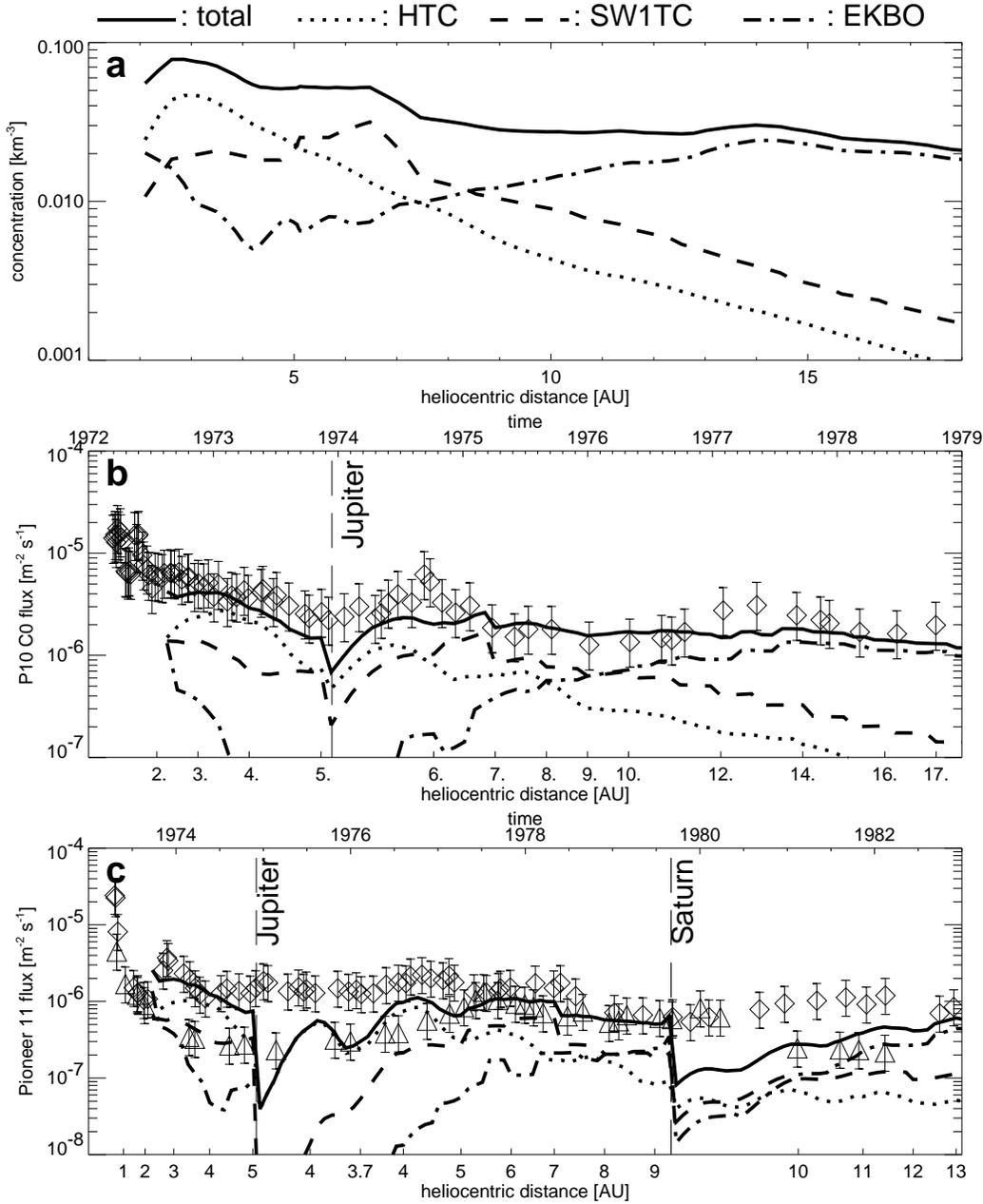}
\caption{\label{fig_profile}\it\small Radial profiles of the
distribution of interplanetary dust in the outer solar system. The
concentration of dust particles from 1P/Halley-type comets (HTC),
29P/Schwassmann-Wachmann-1-type comets (SW1TC), and Edgeworth-Kuiper
Belt Objects (EKBO) that is needed to account for the Pioneer 10
measurements is shown in {\bf a}. The comparison {\bf b} of the
calculated radial flux signatures of the various sources with the
penetration fluxes measured by Pioneer 10 (diamonds, error bars
indicating $1\sigma$ errors) that particles
from HTCs contribute mainly inside Jupiter's orbit, SW1TC particles
between $6$ and $7\:{\rm AU}$, and particles from EKBOs dominate
outside $10\:{\rm AU}$. The profile of the penetration flux of the
Pioneer 11 dust instrument ({\bf c}, diamonds: channel 0 data,
triangles: channel 1 data) is very flat due to the triple passage of
Pioneer 11 through the $4$-to-$5$-AU region.}
\end{figure*}

What is the signature of particles from HTCs, SW1TCs, and EKBOs in the
Pioneer data? The particles' equilibrium distribution in the solar
system is determined by their initial orbit after they have been
released from the source object\footnote{Or equivalently from ${\rm
cm}$-sized fragments that form the source object's trail along its
orbit.}, and by their orbital evolution under PR and solar wind drag,
as well as under gravitational perturbations by the planets. The
effect of the planet's gravity on the grains is strongest when the
orbital period of the planet and the particle have an integer ratio,
that is when the particle is in a mean motion resonance (MMR) with the
planet. An MMR is described by the ratio $p:q$, where $q$ is the
number of orbits the particle completes in the time the planet orbits
the Sun $p$ times. The effect of exterior MMRs, for which $p>q$, as
well as on the spatial distribution and orbits of dust particles in
the solar system has been predicted \citep{jackson89} and observed
\citep{dermott94}. When a particle is in an exterior MMR, it's Sun-ward
motion is temporarily halted, because the energy loss due to PR and
solar wind drag is compensated by the resonant interaction with the
planet's gravity field. But then the eccentricity of the particle's
orbit increases until a close encounter with the resonant or a
neighbouring planet ejects the particle from the resonance. Depending
on the planet's mass and the proximity of other strong perturbers, the
exterior MMRs cause a circumsolar dust ring to form. The equilibrium
distribution is achieved when the dust production by the sources is
equalised by the particle sinks, which are evaporation close to the
Sun and ejection from the solar system by close encounters with the
giant planets, mainly Jupiter and Saturn. Due to the long time scales
of orbital evolution, the equilibrium distribution is reached after
$10^5$ to $10^6$ years\footnote{For dust particles with sizes in the
order of $10\:{\rm
\mu m}$.}. This means that not a single comet, the lifetime of which
is typically $10^3$ to $10^4$ years, but only a whole class of comets
with similar orbital characteristics can sustain a equilibrium
distribution. For particles originating from HTCs, it was found
\citep{liou99} that they mainly occupy $p:1$ MMRs with Jupiter, where
$p$ ranges from $2$ to $12$. When they leave the Jupiter resonances,
they continue Sun-ward until they evaporate. Unlike HTC particles,
dust particles released by SW1TCs are not concentrated in exterior
Jupiter MMRs. This is caused by their unstable initial orbits which
bring them close to Jupiter within the first few centuries after their
release from the parent comet. Jupiter perturbs SW1TC particles out
to Neptune's orbit with the maximum spatial concentration at $5$ to
$6\:{\rm AU}$.  Particles originating from EKBOs approach the planets'
orbits from the outside and consequently are found mainly in the
$2:1$, $3:2$, or $4:3$ resonance with Neptune
\citep{liou99b}. After they are ejected from the exterior Neptune MMRs, they
continue to spiral toward the Jupiter/Saturn region, where $80\%$ of
them are ejected from the solar system by close encounters with one of
the giant planets. The other $20\%$ continue to spiral Sun-ward where
they evaporate at a solar distance that depends on their
composition. Figure \ref{fig_profile}a shows the radial profile of the
spatial particle concentration in the solar system for particles from
HTCs, SW1TCs, and EKBOs.

We have simulated the Pioneer 10 and 11 measurements along the
spacecraft's orbits by calculating the flux of dust particles from a
given source on the target surface of the dust detector at the
spacecraft location, given the spacecraft attitude and velocity
vector, and the local dust concentration and velocity vector. Figures
\ref{fig_profile}a and b show the predicted and the measured dust
fluxes on the Pioneer 10 and 11 instruments, respectively. Because the
average dust production rates of the source objects is unknown, we
treated the the total amount of dust, that is the normalisation of the
radial concentration profile, as a free parameter that was established
by a least-square fit of the predictions to the measured values. On
both spacecraft the penetration flux initially decreased due to the
lower dust concentration at larger heliocentric distances. The peak in
the penetration flux measured by Pioneer 10 end-1974 at $6\:{\rm AU}$
is well explained with penetrations caused by particles from HTCs and
SW1TCs. The peak appears to be even stronger than expected from our
calculations. At heliocentric distances of $7\:{\rm AU}$ and beyond,
the constant penetration flux of $2\times 10^{-6}\:{\rm m}^{-2}\:{\rm
s}^{-1}$ can only be explained if we include a substantial
contribution from EKBO particles. At $18\:{\rm AU}$ the flux of EKBO
particles dominates the other two sources by an order of
magnitude. Because the Pioneer 11 dust instrument did not provide much
data beyond the Jupiter-Saturn region, the signature from EKBO
particles is less dominant. Between Jupiter and Saturn, as well as
between Saturn's orbit and a heliocentric distance of $11\:{\rm AU}$,
the contributions from all three sources are comparable.

\section{Dust Production Rates}
The comparison of the measured fluxes with the calculated radial
profiles gives us a direct determination of the dust particle
production rates. In order to provide the penetration fluxes shown in
figure \ref{fig_profile}a, HTCs have to produce $6\times
10^{11}$ , SW1TCs $3\times 10^{11}$, and EKBOs $2\times 10^{14}$ dust
particles of size $10\:{\rm \mu m}$ and larger per second. The
production rate in terms of dust mass is given by the integral of the
production rate over the grain mass distribution. The integration
covers grain masses from the lower sensitivity limit of the Pioneer 10
instrument of $10^{-9}\:{\rm g}$ to an upper limit of $0.1\:{\rm
g}$. The lower mass limit of HTC grains is $10^{-7}\:{\rm g}$, because
the high eccentricity of the source object and solar radiation
pressure cause them to leave the Solar System if they have smaller
masses. The upper limit is determined by the requirement that the
grains have to be distributed by orbital evolution over a large volume
in order to contribute to the interplanetary dust flux measured by
Pioneer. Only grains with masses of less than $0.1\:{\rm g}$ move away
from their parent body's orbits on times scales shorter than the age
of the Solar System. Assuming a generic collision-type grain mass
distribution \citep{dohnanyi72}, we find dust mass production rates of
$3\times 10^5\:{\rm g}\:{\rm s}^{-1}$ for HTCs, $8\times 10^4\:{\rm
g}\:{\rm s}^{-1}$ for SW1TCs, and $5\times 10^7\:{\rm g}\:{\rm
s}^{-1}$ for EKBOs.

\section{Discussion}
From in situ measurements \citep{mazets87} as well as remote sensing
experiments \citep{thomas91} close to the comet's perihelion it was
found that 1P/Halley's dust production rate during its active phase was
$10^7\:{\rm g}\:{\rm s}^{-1}$. Keeping in mind that comet
1P/Halley has an active period that covers less than $1\%$ of its
orbital period, we find that Halley itself produces on average less
than $10^5\:{\rm g}\:{\rm s}^{-1}$. This means that, unless HTCs have
been much more active in the past, there must be a significant
contribution from other sources, like short period Jupiter family
comets, in order to sustain the dust concentration observed by Pioneer
10 between $2$ and $5\:{\rm AU}$.
The measurements by Pioneer 10 at heliocentric distances larger than
$6\:{\rm AU}$ provide better constraints on the dust production rate
of SW1TCs than on the dust production by HTCs. The high penetration
flux measured between $6$ and $7\:{\rm AU}$ can not be explained with
a contribution from HTCs or short period Jupiter family comets. From
the Pioneer 10 measurements we find that, on average, $8\times
10^4\:{\rm g}\:{\rm s}^{-1}$ of dust have to be generated by
SW1TCs. This is considerably lower than the value of $(6\pm 3)\times
10^5\:{\rm g}\:{\rm s}^{-1}$ for the current dust production rate
found by \cite{fulle92} for 29P/Schwassmann-Wachmann 1
itself. This confirms that, due to the proximity of the strong
perturber Jupiter, the dwell time of SW1TCs in their peculiar orbits
is small compared to their lifetimes. This also means that
29P/Schwassmann-Wachmann 1 itself is able to provide a major fraction
of solar system dust that is currently found between $6$ and $8\:{\rm
AU}$.

Our calculations show that the interplanetary dust environment outside
Saturn is dominated by particles originating from EKBOs, unless there
is an unexpected significant contribution from Centaur objects or
unknown sources. If there were a significant amount of dust from
Centaur objects, its spatial density would decrease steeply with
increasing heliocentric distances due to the high eccentricity of the
Centaurs' orbits. Such a radial distribution would not explain the
nearly constant flux observed by Pioneer 10 outside Saturn's orbit. In
order to fit the Pioneer 10 detections outside $10\:{\rm AU}$, dust
has to be produced in the Edgeworth-Kuiper belt at a rate of $5\times
10^7\:{\rm g}\:{\rm s}^{-1}$. Because we assume an equilibrium dust
distribution, this value represents the average over the typical dust
particle lifetime of $10^7\:{\rm years}$. Estimates of the collisional
dust production \citep{stern96}, that include up to kilometre-sized
fragments, give values of $10^9$ to $10^{11}\:{\rm g}\:{\rm
s}^{-1}$. However, the orbits of these large fragments do not evolve
under PR drag into the $10$ to $18\:{\rm AU}$ region. Translating the
collisional production rate into the mass range between $10^{-9}$ and
$0.1\:{\rm g}$ gives a value between $9\times 10^5$ and $3\times
10^8\:{\rm g}\:{\rm s}^{-1}$, depending on the surface properties of
EKBOs. In addition to the collisional dust production, the production
of particles by impacts of interstellar dust grains onto EKBOs was
found to be between $3\times 10^5$ and $3\times 10^7\:{\rm g}\:{\rm
s}^{-1}$
\citep{yamamoto98}. Thus, the EKBO dust production rate derived from
the Pioneer 10 measurements is on the high side of the source
models, but well within the theoretical uncertainties, which include
the size distribution of Edgeworth-Kuiper belt objects, the impactor
flux, and the source objects' surface properties.

\section{Conclusion}
The discussion above shows that we have been able to identify a set of
observable dust sources for the Pioneer dust measurements. Unlike the
interpretation by \cite{humes80}, we have used a set of 3 dynamical
families of source objects. The sum of these sources provides the
right spatial and local velocity distribution that explains the
penetration fluxes measured by Pioneer. We found the calculated
signature of the source families in the data to be independent, that
is dominant for different heliocentric distances, so that dust
production rates for the individual sources could be derived
separately from the data. Especially the data collected by the
spacecraft outside Saturn's orbit is very valuable, because with
increasing heliocentric distance the number of possible contributors
to the interplanetary dust cloud decreases. The only known source of
interplanetary dust outside Saturn is the Edgeworth-Kuiper belt. This
gives us the opportunity to unambiguously determine the amount of dust
released by the objects of the belt. According to the Pioneer 10
measurements, the density of interplanetary dust generated by the
Edgeworth-Kuiper belt is high enough so that this dust cloud is the
second brightest feature of the solar system when observed from afar
\citep{liou99b}. Thus the Edgeworth-Kuiper belt and the distribution
of dust particles it produces can act as a model for detecting other
planetary systems around mid-age main sequence stars. Interplanetary
dust in the region between Jupiter and Saturn gives us information
about the dynamical properties of this interesting region. Since a
fly-by of Jupiter on 31 December 2000 the Cassini spacecraft is
on-route to Saturn, carrying a highly sensitive dust instrument. It
will provide data on the mass, velocity, and chemical composition of
the smaller sized dust particles.

\acknowledgments
\section{Acknowledgements}
Sadly our dear colleague Herbert A. Zook passed away on 14th March
2001. The support by D. H. Humes in various discussions is gratefully
acknowledged. ML thanks D. P. Hamilton for valuable discussions and
M. Khan for improving the clarity of the manuscript. We also
acknowledge the support by the National Space Science Data Center (NSSDC)
which provided a copy of the original Pioneer data. ML was supported
by the National Research Council while the presented work was
performed.

\bibliography{dust}
\bibliographystyle{apj}

\end{document}